\documentclass[10pt,conference]{IEEEtran}
\IEEEoverridecommandlockouts
\usepackage{cite}
\usepackage{amsmath}
\usepackage{amssymb,amsfonts}
\usepackage{algorithmic}
\usepackage{graphicx}
\graphicspath{{images/}}
\usepackage{textcomp}
\usepackage{xcolor}

\usepackage{lineno}
\usepackage{ulem}
\usepackage{hyperref}
\usepackage{algorithm}
\usepackage{algorithmic}
\usepackage{diagbox}
\usepackage{subfigure}
\usepackage{bbding}
\usepackage{fancyhdr}

\makeatletter
\newcommand{\removelatexerror}{\let\@latex@error\@gobble}
\makeatother

\newcommand{\mbbE}{\mathbb{E}}
\newcommand{\mbbR}{\mathbb{R}}
\newcommand{\mbbC}{\mathbb{C}}

\newcommand{\mbA}{\mathbf{A}}
\newcommand{\mbB}{\mathbf{B}}

\newcommand{\mbG}{\mathbf{G}}
\newcommand{\mbH}{\mathbf{H}}
\newcommand{\mbI}{\mathbf{I}}
\newcommand{\mbL}{\mathbf{L}}
\newcommand{\mbQ}{\mathbf{Q}}

\newcommand{\mbPi}{\mathbf{\Pi}}
\newcommand{\mbXi}{\mathbf{\Xi}}

\newcommand{\mcS}{\mathcal{S}}
\newcommand{\mcU}{\mathcal{U}}
\newcommand{\mcC}{\mathcal{C}}

\newcommand*{\dif}{\mathop{}\!\mathrm{d}}
\newcommand*{\cmac}{C_{\mathrm{MAC}}}

\newtheorem{theorem}{\hspace{-1em}\textbf{Theorem}}

\newcommand{\lbrac}[1]{\left\{#1\right\}}

\def\BibTeX{{\rm B\kern-.05em{\sc i\kern-.025em b}\kern-.08em
    T\kern-.1667em\lower.7ex\hbox{E}\kern-.125emX}}
    
\IEEEoverridecommandlockouts\IEEEpubid{\makebox[\columnwidth]{ 978-1-6654-3540-6/22/\$31.00 \copyright~2022 IEEE \hfill} \hspace{\columnsep}\makebox[\columnwidth]{ }}
    
\begin{document}
%\linenumbers

\title{The Moment Passing Method for Wireless Channel Capacity Estimation
%MoSE: An Accurate and Efficient Algorithm for  Wireless Channel Capacity Estimation Based on Moments and Random Matrix Theory
\thanks{This work was supported by National Natural Science Foundation of China Grant Nos. 11971371 and 11871297, Key Technologies for Coordination and Interoperation of Power Distribution Service Resource 2021YFB2401300, and Fundamental Research Funds for the Central Universities.} 
}

\author{\IEEEauthorblockN{Han Hao\IEEEauthorrefmark{2}, Dandan Jiang\IEEEauthorrefmark{3}, Lu Yang\IEEEauthorrefmark{4}\Envelope, Hao Wu\IEEEauthorrefmark{5}, and Bo Bai\IEEEauthorrefmark{4}}%Xiang Chen\IEEEauthorrefmark{4}, Wei Han\IEEEauthorrefmark{4}, 
  \IEEEauthorblockA{\IEEEauthorrefmark{2}School of Aerospace Engineering, Tsinghua University, Beijing, China}
  \IEEEauthorblockA{\IEEEauthorrefmark{3}School of Mathematics and Statistics, Xi'an Jiaotong University, Shaanxi, China}
  \IEEEauthorblockA{\IEEEauthorrefmark{4}Theory Lab, Central Research Institute, 2012 Labs, Huawei Technologies Co. Ltd., Hong Kong SAR, China}
  \IEEEauthorblockA{\IEEEauthorrefmark{5}Department of Mathematical Sciences, Tsinghua University, Beijing, China}
  \IEEEauthorblockA{Email: haoh19@mails.tsinghua.edu.cn, jiangdd@xjtu.edu.cn,\\
  yanglu87@huawei.com, hwu@tsinghua.edu.cn, baibo8@huawei.com}
 }%chenxiang73@huawei.com, harvey.hanwei@huawei.com,

\maketitle
 
\thispagestyle{fancy}
\fancyhead{}
\lhead{}
\lfoot{\small\copyright~2022 IEEE.  Personal use of this material is permitted.  Permission from IEEE must be obtained for all other uses, in any current or future media, including reprinting/republishing this material for advertising or promotional purposes, creating new collective works, for resale or redistribution to servers or lists, or reuse of any copyrighted component of this work in other works.}
\cfoot{}
\rfoot{}

\begin{abstract}
% Wireless network capacity can be regarded as the most important performance metric for wireless communication systems.
% With the fast development of wireless communication technology, future wireless systems will become more and more complicated.
% As a result, the channel gain matrix will become a large-dimensional random matrix, leading to an extremely high computational cost to obtain the capacity. 
% %Currently, determining the capacity based on the large-dimensional random matrices lacks fast algorithms.
% In this paper, we propose a moment passing method (MPM) to realize the fast and accurate capacity estimation for future ultra-dense wireless systems.
% With this method, we can estimate the capacity under arbitrary distributions of base stations (BSs) and users, and the shape of network areas. 
% The estimation error is below $2\%$, and the time complexity is only quadratic. 
% Moreover, MPM can be applied not only to the conventional multi-user multiple input and multiple output (MU-MIMO) networks, but also to the capacity-centric (C$^2$) networks designed for B5G/6G.
Wireless network capacity can be regarded as the most important performance metric for wireless communication systems. With the fast development of wireless communication technology, future wireless systems will become more and more complicated. As a result, the channel gain matrix will become a large-dimensional random matrix, leading to an extremely high computational cost to obtain the capacity. In this paper, we propose a moment passing method (MPM) to realize the fast and accurate capacity estimation for future ultra-dense wireless systems. It can determine the capacity with quadratic complexity, which is optimal considering that the cost of a single matrix operation is not less than quadratic complexity. Moreover, it has high accuracy. The simulation results show that the estimation error of this method is below 2\%. Finally, our method is highly general, as it is independent of the distributions of BSs and users, and the shape of network areas. More importantly, it can be applied not only to the conventional multi-user multiple input and multiple output (MU-MIMO) networks, but also to the capacity-centric networks designed for B5G/6G.
\end{abstract}


\section{Introduction}\label{section1}
Wireless network capacity can be regarded as the most important performance metric for wireless communication systems.
In 1948, a mathematical model to determine the channel capacity was first given by Dr. Claude E. Shannon\cite{shannon1948mathematical}. 
He proposed that the capacity of a single input and single output channel can be determined according to the formula: $C=W \log (1+\frac{P}{N_0 W})$ \cite{cover2006elements}, where $W$ is the channel bandwidth, $P$ is the signal power and $N_0$ is the power of additive white Gaussian noise (AWGN).
With the fast development of wireless technology, multiple antennas were equipped on both of the transmitter side and the receiver side, in order to improve the capacity. 
The uplink capacity of a multi-user multiple input and multiple output (MU-MIMO) network, which is equivalent to the capacity of a multiple access channel (MAC), can be determined according to \cite{tse2005fundamentals} as:
\begin{equation}\label{cap_MIMO}
    \cmac\!=\!\mbbE \{\log \det (\mbI \!+\! \frac{P}{N_0} \mbH \mbH^*)\}\!=\!\mbbE \sum_{j=1}^t \log(1\!+\!\frac{P}{N_0}\lambda_j),
\end{equation}% david xie p426-428
% is difficult to calculate when large
% start from MIMO
% either large matrices or eigenvalues, both high complexity
% when r and t large tx, rx
% cellular: t=1, 1 moment is enough
% it is important to reduce in B5G
%{\color{red} where $P$ and $N_0$ are the transmitting power and the noise power, respectively.}
where $\mbH \in \mbbC^{t\times r}$ denotes the channel gain matrix, with $t$ and $r$ representing the number of transmit antennas and the number of receive antennas, respectively.
$\mbH^*$ is the Hermitian conjugate of matrix $\mbH$.
% $\frac{P}{N_0} \mbH \mbH^*$ is called the Signal to Noise Ratio (SNR) matrix of the channel, since it represents the SNR of the channel.
$\det (\cdot)$ represents the determinant of a matrix.
$\lambda_1, \cdots \lambda_t$ are the eigenvalues of the matrix $\mbH \mbH^*$.

It is well known that the computation of MAC capacity $\cmac$ is complicated.
%It can be observed from \eqref{cap_MIMO} that determining the MAC capacity $\cmac$ requires matrix operations and eigenvalue derivations. 
%The complexity of conventional methods to determine $\cmac$ are always $O(t^3)$, such as the Cholesky decomposition method, the singular value decomposition (SVD) method, etc.
% Cholesky decomposition or SVD of the SNR matrix, whose complexity are both $O(t^3)$.
Specifically, $\mbH$ will become a large-dimensional random matrix as $t$ and $r$ increase. 
The matrix multiplication $\mbH \mbH^*$ consumes expensive time cost. 
The complexity of direct methods to compute $\cmac$, such as the Cholesky decomposition method (CDM) and the singular value decomposition (SVD) method, reaches $O(t^3)$. 
This is unaffordable for future ultra-dense networks.

In 2021, a capacity-centric (C$^2$) networking architecture designed for future ultra-dense wireless networks was proposed \cite{yang2021future}. 
It can realize both high capacity and low signaling overhead simultaneously.
The C$^2$ networking architecture is organized in the form of multiple non-overlapping clusters, with each cluster operating independently. 
Only base stations (BSs) belonging to the same cluster cooperate with each other, and the interference exists between different clusters, which can be called as the inter-cluster interference. 
Therefore, each cluster is equivalent to a MU-MIMO subnetwork, and a C$^2$ network can be regarded as a wireless network composed of multiple MU-MIMO subnetworks with interference between different subnetworks.
For a C$^2$ network, the average capacity of the $m$-th cluster per BS is given in \cite{yang2021future} as
%{\color{red} Besides of the aforementioned point-to-point channel. How to derive the capacity for a more complicated wireless system, where the interference can not be ignored? Let's focus on a wireless system composed of multiple non-overlapping clusters, where each cluster operates independently, and the interference exist among different clusters, called as inter-cluster interference. Inside each cluster, there are multiple single-antenna users and single-antenna base stations (BSs). Thus, if we ignore the inter-cluster interference, the average capacity of each cluster is equivalent to $\cmac$ \cite{wang2021clustered}.While if the inter-cluster interference can not be ignored, the uplink average  capacity of the $m$-th cluster can be obtained according to \cite{yang2021future} as}
\begin{equation}\label{cap_def_pre}
    C_m \!=\!\mbbE \!\lbrac{ \!\frac{1}{J_m} \!\log \det \left[\mbI \!+\! P(N_0 \mbI \!+\! P \mbPi_m \mbPi_m^*)^{-1}\mbH_m \mbH_m^*\right]\!}\!,
\end{equation}
where $\mbH_m$ and $\mbPi_m$ are the channel gain matrix and the interference matrix of cluster $m$ respectively, which will be specified later.
With a little abuse of notation, we denote the term $P(N_0 \mbI \!+\! P \mbPi_m \mbPi_m^*)^{-1}\mbH_m \mbH_m^*$ as the signal to interference plus noise ratio (SINR) matrix.
%, and thus we call it the SINR matrix for brevity.
$J_m$ denotes the number of BSs\footnote{Here we simply consider the situation of single-antenna BS. For multi-antenna BS, $J_m$ can also denote the total number of antennas.}.
Obviously, \eqref{cap_def_pre} is much more complicated than \eqref{cap_MIMO} since \eqref{cap_def_pre} contains the inter-cluster interference.
It is worth mentioning that in the special case where the number of clusters equals to one, all the BSs cooperate with each other, and no inter-cluster interference exists. 
The uplink channel of such a wireless network is equivalent to a MAC channel. 
That is, the formula \eqref{cap_def_pre} degenerates to \eqref{cap_MIMO}. Therefore, without loss of generality, we will focus on the expression \eqref{cap_def_pre} and develop a fast and accurate numerical method to estimate $C_m$.
%More importantly, for a special case of  C$^2$ where the number of clusters is 1, all the BSs are cooperated, and there is no inter-cluster interference. 
%The uplink capacity of C$^2$ in the ultra-dense networks in the special case can thus be regarded as the capacity of a MAC, and \eqref{cap_def_pre} in this special case becomes equivalent to \eqref{cap_MIMO}.
%Therefore, in the following part of this paper, we will focus on the more general expression \eqref{cap_def_pre}, and design a fast and accurate algorithm to estimate $C_m$.
% This will lead to more difficulty in determining the capacity than the traditional MAC channel.
% In B5G/6G scenarios, where both the BSs and users are densely deployed, the capacity calculation according to \eqref{cap_MIMO} is with high complexity.
% It involves matrix products, inversion of matrices and determinant, all of which have cubic complexity.
% Thus, it is inevitable to explore a fast and accurate method to calculate the capacity for future wireless communication networks.

There are already some existing works to achieve the goal. 
In \cite{yang2021future}, the closed-form expressions of the upper and lower bounds of $C_m$ for uniformly distributed users scenario are derived. 
However, we don’t yet know how to generalize this to more general scenarios. 
Later, the TOSE algorithm \cite{jiang2022tose} was proposed based on the random matrix theory (RMT) \cite{tulino2004random} for fast estimation of $C_m$, but it is less accurate.

%The main purpose of this paper is to design a fast, accurate and general algorithm to determine the average cluster capacity $C_m$ in \eqref{cap_def_pre}. 
% It can be observed from \eqref{cap_def_pre} that the direct calculation of $C_m$ is extremely complicated due to the unavoidable matrix manipulations. 
% Especially, for future ultra-dense scenarios, the computational cost will become unaffordable, since both $\mbH_m$ and $\mbPi_m$ become large-dimensional random matrices. % move to the upper paragraph
%Although the closed-form expressions of an upper bound and a lower bound of $C_m$ for a special scenario of uniformly distributed users are derived in \cite{yang2021future}, the accurate capacity estimation of the capacity $C_m$ for the general scenario is still missing.
%In our previous work \cite{jiang2022tose}, a TOSE algorithm was proposed to estimate $C_m$. 
%Although TOSE is fast, its accuracy is low.
%Can we find an algorithm, which is fast, accurate, and general enough to estimate the capacity?
% The TOSE algorithm is fast enough, but its accuracy should be better optimized.
% What's more, we want to develop an algorithm that can determine the capacity of both MAC and C$^2$ architecture for future wireless networks.

% fast, accurate, general
% has great generality: first study 2, when interference is 0, 2 becomes 1.

In this paper, we propose a Moment Passing Method (MPM) to estimate the capacity of ultra-dense wireless networks.
Inspired by the Mar\v{c}enko-Pastur (MP) law in RMT, we propose a numerical approach to approximate the limiting spectral distribution (LSD) of the SINR matrix. 
By utilizing the characteristics of the random matrix, the high-complexity procedures including the matrix multiplication and the eigenvalue computation can be avoided. 
Therefore, our MPM is a low-complexity method.
Different from the TOSE method \cite{jiang2022tose}, we directly estimate the LSD of the SINR matrix, avoiding an intermediate step of SINR matrix approximation in TOSE, and thus our MPM has higher accuracy.
It should be further emphasized that the main novelty of our MPM is to combine the Stieltjes transform and the Laurent expansion to obtain the moment information of the LSD. 
This is the reason why we call this method Moment Passing Method (MPM).
In the numerical simulations, we will show that the computational efficiency of MPM is almost identical to TOSE, but the numerical error is reduced by at least an order of magnitude compared with TOSE.
In addition, MPM has superior generality since it is independent of the distributions of BSs and users, and the shapes of network areas.
Finally, MPM is applicable not only to the conventional MU-MIMO networks, but also to the C$^2$ networks designed for B5G/6G.

\section{System Model}\label{section2}

% \subsection{System Model and Capacity Approximation}\label{section21}

Consider a wireless network with $J$ single-antenna BSs $\mcS=\{s_1, s_2, \dots s_J \}$ and $K$ single-antenna users $\mcU=\{u_1, u_2,\dots,u_K\}$. 
The whole network is organized in the form of $M$ non-overlapping clusters $\mcS\bigcup \mcU =\bigcup_{m=1}^M \mcC_m$ \cite{yang2021future}. 
Here, $\mcC_m$ represents the set of BSs and users belonging to the $m$-th cluster.
The set of the BSs in $\mcC_m$ is denoted by $\mcS_m=\mcS\bigcap \mcC_m$, and similarly, the set of the users in $\mcC_m$ is  $\mcU_m=\mcU\bigcap \mcC_m$.
And we denote $J_m=| \mcS_m|$ and $K_m=| \mcU_m|$ the number of BSs and users in $\mcC_m$, respectively. 
%Moreover, we use $J_m=| \mcS_m|$ and $K_m=| \mcU_m|$ to denote the number of BSs and users in $\mcC_m$, respectively. 
To reflect the property of ultra-dense networks, we assume $J_m$ and $K_m$ approach infinity \cite{yang2021future}.

For the $m$-th cluster, we define the channel gain between the BS $s_j \in \mcS_m$ and the user $u_k \in \mcU$ as $h_{mjk}=l_{mjk} g_{mjk}$, where $g_{mjk}\sim\mathcal{CN}(0,1)$ is the small-scale fading and 
\begin{equation}\label{l_def}
	l_{mjk}=\left\{\begin{array}{ll}
		d_{mjk}^{-1.75}, & d_{mjk}>d_1, \\
		d_1^{-0.75} d_{mjk}^{-1}, & d_0<d_{mjk}\le d_1, \\
		d_1^{-0.75}d_0^{-1}, & d_{mjk}\le d_0
	\end{array}\right.
\end{equation}
is the large-scale fading \cite{yang2021future}.
Here $d_{mjk}$ represents the Euclidean distance between the BS $s_j \in \mcS_m$ and the user $u_k$. 
The parameters $d_0$ and $d_1$ can be regarded as the near field threshold and far field threshold, respectively.

Thus, we can define the large-scale fading matrix $\mbL_m\in\mbbR^{J_m\times K_m}$ and the small-scaling fading matrix $\mbG_m\in\mbbC^{J_m\times K_m}$, with their $(j,k)$-th entry given by
\begin{equation*}
	[\mbL_m]_{jk}=l_{mjk}, \quad
	[\mbG_m]_{jk}=g_{mjk}.
\end{equation*}
The channel gain matrix $\mbH_m$ in \eqref{cap_def_pre} can thus be defined as
\begin{equation}
	\mbH_m=\mbL_m\circ\mbG_m,
\end{equation}
in which $\circ$ denotes the Hadamard product. 
The interference matrix $\mbPi_m$ can be similarly defined since it is also a Hadamard product between a large-scale fading matrix and a small-scale fading matrix. 
Details of deriving $\mbPi_m$ can be found in \cite{deng2022cgn}, and thus we do not repeat here due to space limitation. 
Define
\begin{equation}
    \mbXi_m=N_0 \mbI + P \mbPi_m \mbPi_m^*,
\end{equation}
as the noise-plus-interference matrix, with $\mbXi_m\in \mbbC^{J_m \times J_m}$. It converges to a positive definite diagonal matrix as $J_m$ and $K-K_m$ approach infinity \cite{deng2022cgn}, which is 
%Based on Lemma 1 in \cite{deng2022cgn}, $\mbXi_m$ converges to a positive definite diagonal matrix as $J_m$ and $K-K_m$ approach infinity, which is
\begin{equation}\label{sjj_def}
	\mbXi_m=\textrm{diag}((N_0+P \xi_{11}^m),\dots,(N_0+P \xi_{J_m J_m}^m)),
\end{equation}
with $\xi_{jj}^m=\sum_{ u_k\notin \mcU_m} l_{mjk}^2$. 
Substituting it into \eqref{cap_def_pre}, we have
%By substituting \eqref{sjj_def} into \eqref{cap_def_pre}, we have
\begin{align}
	C_m &=\mbbE\left\{ \frac{1}{J_m}\log \det \left(\mbI + P\mbXi_m^{-1/2} \mbH_m \mbH_m^* \mbXi_m^{-1/2}\right) \right\} \nonumber \\
	&=\mbbE\left\{ \frac{1}{J_m}\log \det \left[\mbI + (\mbQ_m\circ \mbG_m)(\mbQ_m\circ \mbG_m)^*\right] \right\},\label{cap1}
\end{align}
in which $\mbQ_m=P^{1/2}\mbXi_m^{-1/2}\mbL_m$. 
%In the following, we will focus on the computation of $C_m$. 

%Note that there exists an special case. 
%If there is only one cluster in the C$^2$ architecture (i.e., $M=1$), there is no inter-cluster interference, and thus $\mbPi$ is zero.
%Then we have $\mbXi_m= N_0 \mbI$, and $C_m$ becomes
%\begin{equation}\label{cap_mac}
%    C_m\!\Big\arrowvert_{M=1}\!=\!C_{\mathrm{MAC}}\!=\!\mbbE\!\left\{\! \frac{1}{J_m}\!\log \det \!\left[\mbI + \frac{P}{N_0}(\mbL_m\circ \mbG_m)(\mbL_m\circ \mbG_m)^*\right] \!\right\}.
%\end{equation}

%In this special case, the capacity of the C$^2$ network becomes equivalent to the capacity of a MAC channel, since all the BSs are cooperated and there is no inter-cluster interference. 
%Therefore, \eqref{cap1} is more general, and \eqref{cap_mac} is a special case of \eqref{cap1}.

In the following section, we will focus on \eqref{cap1} to develop the fast and accurate Moment Passing Method for the capacity estimation. 
As we have mentioned, $\cmac$ in \eqref{cap_MIMO} is a special case of $C_m$ \eqref{cap1}. 
Therefore, the method developed in this paper is obviously applicable to \eqref{cap_MIMO}. 
\section{The Moment Passing Method}

In this section, we will introduce our MPM to calculate the capacity based on \eqref{cap1}.
% It can be written as:
%In this section, we propose a MPM based algorithm to estimate the capacity, based on calculating the moments of the limiting spectral distribution (LSD) of the random matrix. 
First, we transform the original problem of calculating matrix determinant \eqref{cap1} into the problem of computing the product of all matrix eigenvalues, as
\begin{equation} \label{eqn:cap_equ}
    C_m=\mbbE\!\left[ \frac{1}{J_m}\log \det (\mbI \!+\! \mbA_m) \right]=\frac{1}{J_m}\mbbE\sum_{j=1}^{J_m}\log (1\!+\!\lambda_j).
\end{equation}
Here the SINR matrix $\mbA_m=\mbB_m \mbB_m^*\in\mbbC^{J_m \times J_m}$ with $\mbB_m=\mbQ_m \circ \mbG_m$, and $\lambda_j\;(j=1,2,\cdots,J)$ gives all the eigenvalues of the SINR matrix $\mbA_m$. 
As we mentioned before, it is very expensive to solve all the eigenvalues of a large matrix, especially for $J_m\to\infty$. 
Moreover, we need to generate the SINR matrix $\mbA_m$ many times to obtain the expectation value in \eqref{eqn:cap_equ}. 
% {\color{red}Fortunately, for the aforementioned two problems, RMT provides us an approximate solution \cite{bai2010spectral}, it writes} 
% {\color{blue}Fortunately, when $J_m$ approaches infinity, \eqref{eqn:cap_equ} can be approximated as}
%based on which we have %\cite{jiang2022tose}
Fortunately, based on RMT \cite{bai2010spectral}, we can approximate \eqref{eqn:cap_equ} using the following integral form
\begin{equation}\label{cap_int}
    C_m\approx\int_{a}^{b}\log(1+x)f(x) \dif x,
\end{equation}
in which $f(x)$ is the LSD of $\mbA_m$. 
The lower and upper limits of the integral interval $[a,b]$ correspond to the minimum and maximum eigenvalues of the SINR matrix $\mbA_m$, respectively. 
Here, we introduce a hyper-parameter $\eta$, which can be understood as an approximation of the ratio between the minimum and maximum eigenvalues.

%{\color{red}In order to make use of the MP-law in RMT \cite{bai2010spectral}, we define  $\widetilde{\mbA}_m=\mbG_m \mbG_m^*$. 
%It can be interpreted as a special case of $\mbA_m$ where the matrix $\mbQ_m$ is an all-ones matrix.
%Then, according to the MP-law, the LSD of $\tilde{\mbA}_m$ can be given by}
According to the MP-law in RMT \cite{bai2010spectral}, the LSD of the matrix $\widetilde{\mbA}_m=\mbG_m \mbG_m^*$, i.e. $\mbQ_m$ is an all-ones matrix, can be approximately written as\footnote{In the MP-law of RMT, we actually consider the LSD of the matrix $\frac{1}{K_m}\mbG_m \mbG_m^*$. The discussion here is not rigorous enough, but it is numerically reasonable and easier for writing.}
\begin{subequations}
\begin{align}
    &\widetilde{f}(x)\!=\!\frac{\beta }{2\pi K_m x}\sqrt{(b\!-\!x)(x\!-\!a)}, \!&\!\beta\!\ge \!1,\label{MP1}\\
    &\widetilde{f}(x)\!=\!(1\!-\!\beta)\delta(x)\!+\!\frac{\beta }{2\pi K_m x}\sqrt{(b\!-\!x)(x\!-\!a)}, \!&\!\beta\!<\!1.\label{MP2}
\end{align}
\end{subequations}
Note that the parameter $\beta=K_m/J_m$ in the above formulas represents the width-length ratio of the matrix $\mbG_m$. 
For $\beta<1$, the matrix $\widetilde{\mbA}_m$ is not full rank. 
Therefore, \eqref{MP2} introduces the delta function $\delta(x)$ to characterize the distribution of zero eigenvalues. 

% In the conditions where $\beta<1$, $\widetilde{\mbA}_m$ is not of full rank, so there are many zero eigenvalues.
% We use the delta function to represent these eigenvalues.

%{\color{red}Then, let's move to the general case, to investigate the LSD of the SINR matrix $\mbA_m$. 
%For such general case, there is no analytical expressions to quantify the LSD of $\mbA_m$ currently. 
%However, we choose the way of making a hypothesis that the LSD of $\mbA_m$ can be approximated by adding a $N$-order polynomial correction to the standard MP-law, i.e. }
For the LSD of the SINR matrix $\mbA_m$, there is currently no result in its analytic form. 
In this work, we hypothesize it can be approximated by a $N-$th order polynomial correction to the standard MP-law, i.e. 
\begin{subequations}\label{app_LSD}
\begin{align}
    &f(x)\!=\!\frac{\beta \!\sqrt{(b-x)(x-a)}}{2\pi K_m x}\sum_{k=0}^{N}\alpha_k x^k, &\beta\!\ge\! 1,\\
    &f(x)\!=\!(1\!-\!\beta)\delta(x)\!+\!\frac{\beta \!\sqrt{(b-x)(x-a)}}{2\pi K_m x}\sum_{k=0}^{N}\alpha_k x^k, &\beta\!<\!1.
\end{align}
\end{subequations}
The rationality of the modification is that elements in $\mbQ_m$ are monotonically related to the distance between BSs and users in the network. 
And the difference between the elements in the matrix $\mbQ_m$ are limited. 
In the numerical experiments, we will show that the approximation to LSD gives a good estimation of capacity. 
As $N$ increases, the accuracy of capacity estimation is also improved. 
According to the numerical results, $N=3$ is good enough to estimate the capacity.

%The distribution of entries in $\mbQ_m$ is continuous since they are related to the distances between BSs and users in the network.
%Then we assume that the LSD of $\mbA_m$ is an amendment of the standard MP-law.
%We use the following function to approximate the LSD:

%where $N$ is a finite positive integer. 
%In the numerical experiments in Section \ref{section4}, we find that the result of $N=3$ is good enough.

%$ b, a $ are the largest and smallest non-zero eigenvalues of $\mbA_m$.
%We define $\eta=a/b$ here.

% To approximate $\log(1+x)$ as a polynomial $(c_0+c_1 x +c_2 x^2 + c_3 x^3)$ between $a$ and $b$, we use the measure

% \begin{equation}
%     \arg \min_{c_0, c_1, c_2, c_3}\int_a^b [\log(x)-(c_0+c_1 x +c_2 x^2 + c_3 x^3)]^2 \dif x,
% \end{equation}
% which is a quadratic function of $c_i$, so we can solve it with simple linear calculations.

% Then the average cluster capacity can be written as:

% \begin{equation}
% \begin{split}
%     C_m&=\int_a^b \log(1+x) f(x) \dif x \\
%     &\simeq \int_a^b (c_0+c_1 x +c_2 x^2 + c_3 x^3) f(x) \dif x\\
%     &=c_0 + c_1 \phi_1 + c_2 \phi_2 + c_3 \phi_3,
% \end{split}
% \end{equation}

% where $\phi_i$ is the $i$-th moment of $f(x)$.

% To determine the coefficients $\alpha_k$, we need the moments of the LSD, which can be determined with the following theorem:

Next, we compute the polynomial coefficients $\alpha_k$ in \eqref{app_LSD} by using the moments 
\begin{equation} \label{eqn:LSD_moments}
    \phi_j=\int_a^b x^jf(x)\dif x, \quad j=0,1,\cdots,N.
\end{equation}
of the LSD of the SINR matix $\mbA_m$. 
By substituting \eqref{app_LSD} into \eqref{eqn:LSD_moments}, we have
\begin{equation}\label{fit_f}
    \sum_{k=0}^{N}c_{j+k}\alpha_k = \phi_j,\quad j=0,1,\cdots,N.
\end{equation}
The parameters $c_i$ in the above formula can be pre-calculated as follows
\begin{equation*}
    c_i=\int_a^b x^i \frac{\beta \sqrt{(b-x)(x-a)}}{2\pi K_m x}\dif x.
\end{equation*}
%For the polynomial coefficient $\alpha_k$ in \eqref{app_LSD}, we can 
%The key of solving the problem becomes determining the coefficients $\alpha_k$.
%In fact, we can easily calculate them with the relationships between the moments and the coefficients of weighted polynomials.
%Plugging \eqref{app_LSD} into \eqref{cap_int}, we have
%where $c_i=\int_a^b x^i \frac{\beta \sqrt{(b-x)(x-a)}}{2\pi x}\dif x$, $\phi_j$ is the $j$-th moment of $f(x)$, and $\phi_0=\min(1,\beta)$.

Finally, we need to get the moments $\phi_j$ of the LSD $f(x)$. 
They can be obtained by the Stieltjes transform \cite{wall1948analytic} of the LSD $f(x)$, combined with the Laurent expansion \cite{rudin1966real}. 
This is the crucial point of our work, which we also illustrate with the following theorem. 
Since the information is transferred through the moments $\phi_j$, we name this method the Moment Passing Method (MPM).

\vspace{1em}
\begin{theorem}\label{theorem1}
Consider the given matrix $\mbQ=[q_{ij}] \in \mbbR^{p\times n}$ and the random matrix $\mbG\in \mbbC^{p\times n}$, whose elements satisfy complex Gaussian distribution, the moments of the LSD $f(x)$ of the matrix $\mbA=(\mbQ \circ \mbG)(\mbQ \circ \mbG)^*$ are as follows:
\begin{align}
    \phi_0=& 1, \label{moment0}\\
	\phi_1=&\frac{1}{pn}\!\sum_{i=1}^p \sum_{k=1}^n \theta_{ik},\label{moment1}\\
	\phi_2=&\frac{1}{pn^2}\!\sum_{i=1}^p \!\left[\!(\sum_{k=1}^n \theta_{ik})^2 \!+\! \sum_{k=1}^n \theta_{ik}(\sum_{j=1}^p \theta_{jk})\!\right],\label{moment2}\\ 
	\phi_3=&\frac{1}{pn^3}\!\sum_{i=1}^p \!\left[\!(\sum_{k=1}^n \theta_{ik})^3 \!+\! 2(\sum_{k=1}^n \theta_{ik})(\sum_{k=1}^n \theta_{ik}(\sum_{j=1}^p \theta_{jk}))\!\right]\nonumber\\
	+&\frac{1}{pn^3}\!\sum_{i=1}^p \sum_{k=1}^n \theta_{ik}\!\left[\!(\sum_{j=1}^p \theta_{jk})^2 \!-\!(\sum_{j=1}^p \theta_{jk} (\sum_{l=1}^n \theta_{jl}))\!\right]\!,\label{moment3}
% 	\phi_4=&\frac{1}{pn^4}\!\sum_{i=1}^p \!\left[ \!(\sum_{k=1}^n \theta_{ik})^4 \!+\! (\sum_{k=1}^n \theta_{ik}(\sum_{j=1}^p \theta_{jk}))^2\right]\nonumber\\
% 	+&\frac{1}{pn^4}\!\sum_{i=1}^p \sum_{k=1}^n\!\left[\! (\sum_{j=1}^p \theta_{jk})^3 \!+\! 2(\sum_{j=1}^p \theta_{jk})(\sum_{j=1}^p \theta_{jk} (\sum_{l=1}^n \theta_{jl}))\!\right]\nonumber\\
% 	+&\frac{1}{pn^4}\!\sum_{i=1}^p \sum_{k=1}^n \sum_{j=1}^p\!\theta_{ik}\!\left[\! (\sum_{l=1}^n \theta_{jl})^2 \!-\! \sum_{l=1}^n \theta_{jl}(\sum_{m=1}^p \theta_{ml}) \!\right]\nonumber\\
% 	+&\frac{2}{pn^4}\!\sum_{i=1}^p(\sum_{k=1}^n \theta_{ik})\!\sum_{k=1}^n \theta_{ik}\!\left[\! (\sum_{j=1}^p \theta_{jk})^2 \!+\! (\sum_{j=1}^p \theta_{jk} (\sum_{l=1}^n \theta_{jl})) \!\right]\nonumber\\
% 	+&\frac{3}{pn^4}\!\sum_{i=1}^p(\sum_{k=1}^n \theta_{ik})^2 \left[\sum_{k=1}^n \theta_{ik}(\sum_{j=1}^p \theta_{jk})\right], \label{moment4}
\end{align}
with $\theta_{ij}=n q_{ij}^2$.
\end{theorem}
\vspace{1em}

\begin{IEEEproof}
According to \cite{hachem2007deterministic}, the Stieltjes transform of $f(x)$ satisfies
\begin{align}
    s(z)&=\frac{1}{p}\sum_{i=1}^p \psi_i,\\
    \psi_i (z) &= -\left[z(1+\frac{1}{n}\sum_{j=1}^n \theta_{ij}\Tilde{\psi}_j (z))\right]^{-1}, 1\le i \le p,\label{psi_def}\\
    \tilde{\psi}_j (z) &= -\left[z(1+\frac{1}{n}\sum_{i=1}^p \theta_{ij}\psi_i (z))\right]^{-1}, 1\le j \le n.\label{tilde_psi_def}
\end{align}

When $|z|$ is larger than the maximum eigenvalue of $\mbA$, the Laurent expansion of $s(z)$ \cite{bai2010spectral} is
\begin{equation}
    s(z)=-\sum_{k=1}^\infty \frac{\phi_{k-1}}{z^k},
\end{equation}
%where $\phi_k=\mbbE (X^{k})=\int_a^b x^k f_\mbA(x) \dif x$.
Define $u=1/z$, we have
\begin{equation}\label{s_expansion}
    s_\mbA(1/u)=-\phi_0 u-\phi_1 u^2 -\phi_2 u^3 -\phi_3 u^4 +o(u^4).
\end{equation}

Take the Taylor expansion of $\psi_i$ and $\tilde{\psi}_j$, we have
\begin{align}
    \psi_i=a_i u + b_i u^2 + c_i u^3 + d_i u^4 +o(u^4), \label{psi_exp}\\
    \tilde{\psi}_j=\tilde{a}_j u + \tilde{b}_j u^2 + \tilde{c}_j u^3 + \tilde{d}_j u^4 +o(u^4).\label{tilde_psi_exp}
    % s_\mbA\!=\!\sum_{i=1}^p a_i u +\!\sum_{i=1}^p b_i u^2 +\!\sum_{i=1}^p c_i u^3 +\!\sum_{i=1}^p d_i u^4 +o(u^4).
\end{align}

Plugging \eqref{psi_exp}-\eqref{tilde_psi_exp} into \eqref{psi_def}-\eqref{tilde_psi_def}, we obtain
\begin{equation}
\begin{cases}
    a_i=&-1,\\
    \Tilde{a}_j=&-1,\\
    b_i=&\frac{1}{n}\sum_{j=1}^n \theta_{ij}\Tilde{a}_j,\\
    \tilde{b}_j=&\frac{1}{n}\sum_{i=1}^p \theta_{ij}a_j,\\
    c_i=&\frac{1}{n}\sum_{j=1}^n \theta_{ij}\Tilde{b}_j - (\frac{1}{n}\sum_{j=1}^n \theta_{ij}\Tilde{a}_j)^2,\\
    \tilde{c}_i=&\frac{1}{n}\sum_{i=1}^p \theta_{ij}b_i - (\frac{1}{n}\sum_{i=1}^p \theta_{ij}a_i)^2,\\
    d_i=&(\frac{1}{n}\sum_{j=1}^n \theta_{ij}\Tilde{c}_j) + (\frac{1}{n}\sum_{j=1}^n \theta_{ij}\Tilde{a}_j)^3 \\
    &- 2(\frac{1}{n}\sum_{j=1}^n \theta_{ij}\Tilde{a}_j)(\frac{1}{n}\sum_{j=1}^n \theta_{ij}\Tilde{b}_j),\\
    \tilde{d}_j=&(\frac{1}{n}\sum_{i=1}^p \theta_{ij}c_i) + (\frac{1}{n}\sum_{i=1}^p \theta_{ij}a_i)^3 \\
    &- 2(\frac{1}{n}\sum_{i=1}^p \theta_{ij}a_i)(\frac{1}{n}\sum_{i=1}^p \theta_{ij}b_i),
    % e_i=&(\frac{1}{n}\sum_{j=1}^n \theta_{ij}\Tilde{d}_j) - (\frac{1}{n}\sum_{j=1}^n \theta_{ij}\Tilde{a}_j)^4 \\
    % &- (\frac{1}{n}\sum_{j=1}^n \theta_{ij}\Tilde{b}_j)^2\\
    % &- 2(\frac{1}{n}\sum_{j=1}^n \theta_{ij}\Tilde{a}_j) (\frac{1}{n}\sum_{j=1}^n \theta_{ij}\Tilde{c}_j)\\
    % &+ 3(\frac{1}{n}\sum_{j=1}^n \theta_{ij}\Tilde{a}_j)^2(\frac{1}{n}\sum_{j=1}^n \theta_{ij}\Tilde{b}_j),\\
    % \tilde{e}_j=&(\frac{1}{n}\sum_{j=1}^n \theta_{ij}d_i) - (\frac{1}{n}\sum_{j=1}^n \theta_{ij}a_i)^4 \\
    % &- (\frac{1}{n}\sum_{j=1}^n \theta_{ij}b_i)^2\\
    % &- 2(\frac{1}{n}\sum_{j=1}^n \theta_{ij}a_i) (\frac{1}{n}\sum_{j=1}^n \theta_{ij}c_i)\\
    % &+ 3(\frac{1}{n}\sum_{j=1}^n \theta_{ij}a_i)^2(\frac{1}{n}\sum_{j=1}^n \theta_{ij}b_i).
\end{cases}
\end{equation}

Comparing it with \eqref{s_expansion}, we can get the expressions of moments in \eqref{moment0}-\eqref{moment3}.
\end{IEEEproof}

It should be emphasized that higher-order moments can also be obtained using the techniques above. 
However, the form of higher-order moments is too complicated, and we will not show them here. 
Based on the above discussions, we developed the Moment Passing Method, and the pseudo-code is presented in the following Algorithm.

% The following algorithm shows the main procedure of our moment passing method.
% Our idea of estimating the capacity can be concluded by the following algorithm:

\begin{figure}[htbp] 
\vspace{-1em}
\renewcommand{\algorithmicrequire}{\textbf{Input:}}
\renewcommand{\algorithmicensure}{\textbf{Output:}}
\removelatexerror
\begin{algorithm}[H]
	\caption{The Moment Passing Method}
	\begin{algorithmic}[1]
		\REQUIRE Distance $d_{mjk}$, hyper-parameter $\eta$ %The distance $d_{mjk}$ between each BS $s_j$ in $\mcS_m$ and each user $u_k$, $\eta$
		\ENSURE Estimation of $C_m$.
		\STATE Calculate the matrix $\mbL_m, \mbXi_m$ and $\mbQ_m$ with \eqref{l_def} and \eqref{sjj_def}. \label{al0}
		\STATE Calculate the moments of LSD with \eqref{moment0}-\eqref{moment3}. \label{al1}
		\STATE Generate $\mbG_m$, calculate $\mbB_m\!=\!\mbQ_m \circ \mbG_m$.\label{al2}
		\STATE Compute the maximum eigenvalue $b$ of $\mbA_m=\mbB_m \mbB_m^*$, and calculate $a=\eta b$.\label{al3}
		\STATE Use \eqref{cap_int}, \eqref{app_LSD} and \eqref{fit_f} to compute $C_m$ with numerical integration.\label{al4}
		%\STATE Compute $C_m$ with the numerical integration in \eqref{cap_int}.\label{al5} 
		
	\end{algorithmic}
	\label{tose_algorithm}
\end{algorithm}
\vspace{-1em}
\end{figure}

% \begin{equation}\label{fit_matrix}
% \begin{bmatrix}
% \alpha_0 \\ \alpha_1 \\ \alpha_2 \\ \alpha_3 
% \end{bmatrix} = \begin{bmatrix}
% c_0 & c_1 & c_2 & c_3 \\
% c_1 & c_2 & c_3 & c_4 \\
% c_2 & c_3 & c_4 & c_5 \\
% c_3 & c_4 & c_5 & c_6 \\
% \end{bmatrix}
% \begin{bmatrix}
% 1 \\ \phi_1 \\ \phi_2 \\ \phi_3
% \end{bmatrix},
% \end{equation}
% based on which we can get $\alpha_0,\alpha_1,\alpha_2,\alpha_3$ by solving linear equations.

% \subsection{Complexity Analysis}
Here, we use the power method to compute the maximum eigenvalue. In the iterative step, we can use $ x^{(n+1)}=\mbB_m(\mbB_m^* x^{(n)}) $ instead of $ x^{(n+1)}=\mbA_m x^{(n)} $.
This avoids the matrix multiplication operation $\mbA_m=\mbB_m \mbB_m^*$, and thus keeps the overall computational cost of $O(J_m^2)$. %\footnote{In the abstract and conclusion, we replace $J_m$ with $n$. That is, the overall computational complexity is $O(n^2)$, which is more friendly to the readers.}.

Another thing worth noting is that we do not directly calculate the minimum eigenvalue of the matrix. 
If the inverse power method is applied, it will result in the computational cost of $O(J_m^3)$. In fact, we introduce a hyper-parameter $\eta$ to approximate the ratio between the minimum and maximum eigenvalues. According to the numerical simulations, the error of capacity estimation is not sensitive to the selection of hyper-parameter. Not only that, the computation of the non-zero minimum eigenvalue is extremely difficult for $\beta<1$. Therefore, we directly take $\eta=4\times10^{-3}$ in the numerical simulation. This not only ensures numerical accuracy but also avoids excessively expensive computational costs.

\section{Performance Evaluation}\label{section4}
In this section, we demonstrate the high accuracy, efficiency, and generality of our MPM for the wireless capacity estimation through simulations.

\begin{figure}[htbp]
\subfigure{
\begin{minipage}[t]{0.25\linewidth}
\centering
\includegraphics[width=9cm]{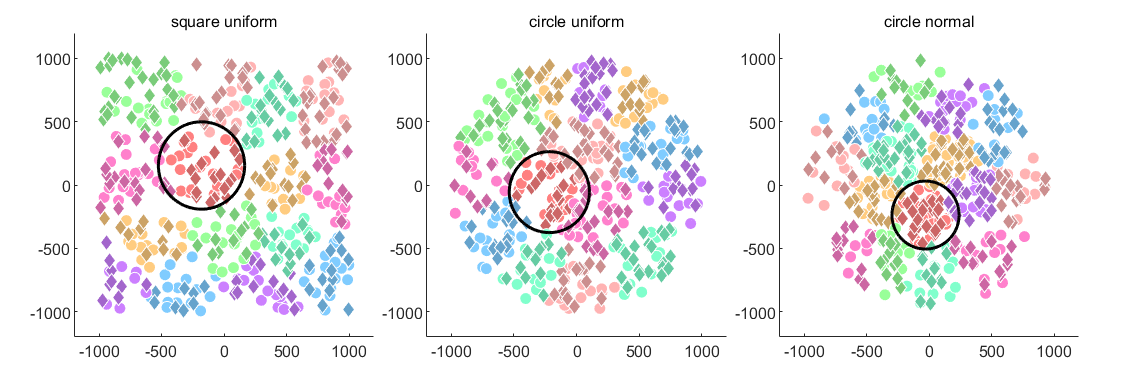}
%\caption{fig1}
\end{minipage}%
}%

\subfigure{
\begin{minipage}[t]{0.25\linewidth}
\centering
\includegraphics[width=9cm]{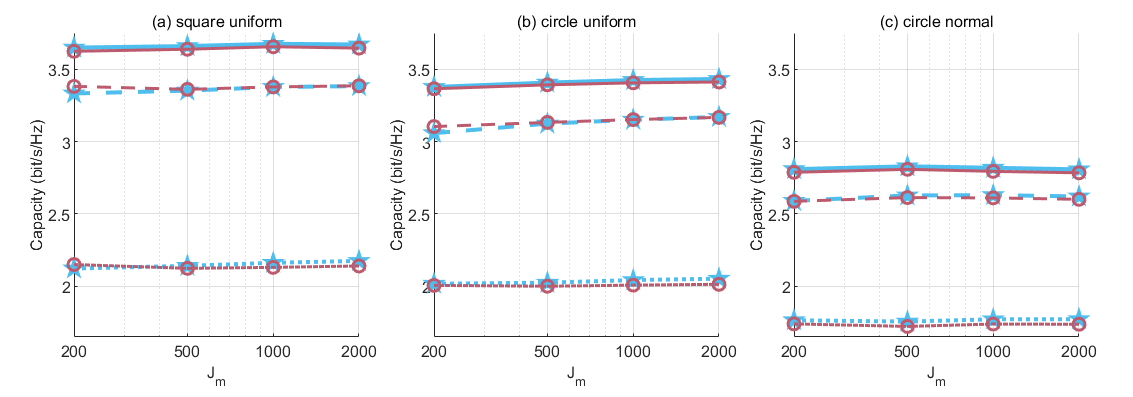}
%\caption{fig2}
\end{minipage}%
}%
\caption{Top row: illustration of C$^2$ networks with three different settings. Different colors represent different clusters. Brighter dots represent BSs, and darker diamonds represent users. Bottom row: the capacity calculated by CDM (blue pentagram) and MPM (red circle). There are three different users-to-BSs ratios, which correspond to the dotted line $\beta=8$, the dashed line $\beta=2$, and the solid line $\beta=0.5$, respectively.}
\label{cap_with}
\vspace{-1.5em}
\end{figure}

\begin{figure}[htbp]
\subfigure{
\begin{minipage}[t]{0.25\linewidth}
\centering
\includegraphics[width=9cm]{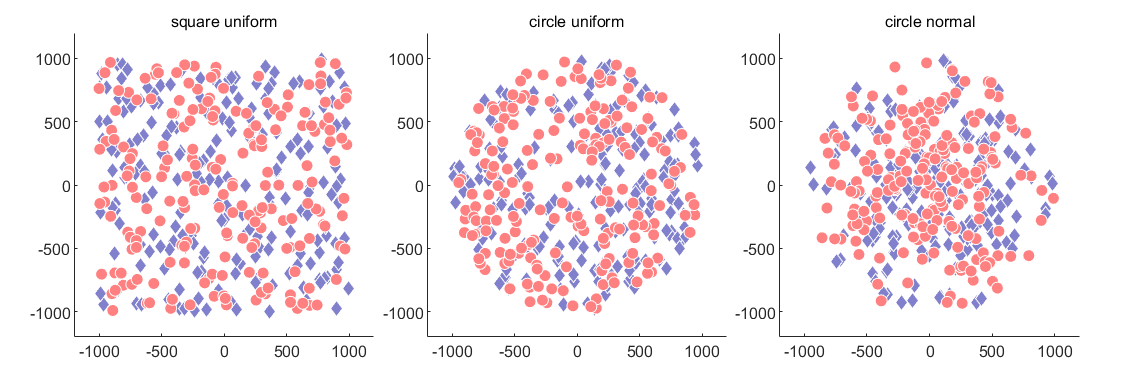}
%\caption{fig1}
\end{minipage}%
}%

\subfigure{
\begin{minipage}[t]{0.25\linewidth}
\centering
\includegraphics[width=9cm]{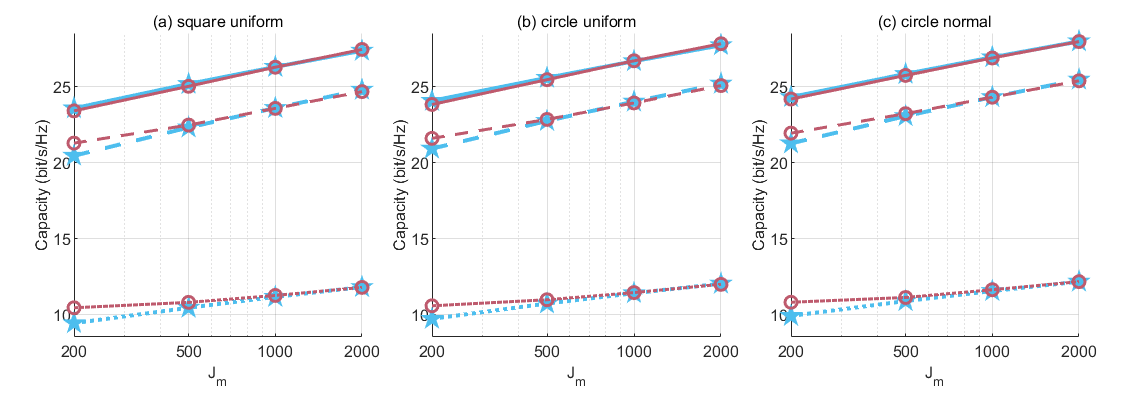}
%\caption{fig2}
\end{minipage}%
}%
\caption{Top row: illustration of MU-MIMO networks, with three different settings. red dots represent BSs, and blue diamonds represent users. Bottom row: the capacity calculated by CDM (blue pentagram) and MPM (red circle). There are three different users-to-BSs ratios, which correspond to the dotted line $\beta=8$, the dashed line $\beta=2$, and the solid line $\beta=0.5$, respectively.}
\label{cap_without}
%\vspace{-1.5em}
\end{figure}

Consider an ultra-dense network, where the network nodes (BSs and users) are deployed according to certain distributions. 
To reveal the generality of our MPM, we simulate two kinds of network areas: circle and square; and two kinds of network node distributions: uniform and normal.
We use $D$ to denote the network diameter for the circle case, and the network side length for the square case.
The whole network is decomposed into $M$ non-overlapping clusters based on the K-means algorithm \cite{lloyd1982least}, as illustrated in Fig. \ref{cap_with}. 
We focused on the capacity of the cluster closest to the center of the entire network, which is highlighted in Fig. \ref{cap_with} by a black circle. 
The simulation results on capacity analysis for other clusters are similar, so we omit them due to space limitation.
We also simulated the special case of C$^2$ in Fig. \ref{cap_without}, where all the BSs cooperate and the number of clusters is one.
This special case of C$^2$ is equivalent to a MU-MIMO network as discussed before.
%Then we choose the cluster closest to the center of the whole network to investigate its capacity.
%Other clusters will have the same properties, and thus we omit them due to space limitation.
% In order to show the generality of our algorithm, we will consider two shapes of network areas: circle and square, and two distributions of network nodes: uniform and normal. 
%Moreover, not only the C$^2$ architecture is simulated, but also its special case with a single cluster (which is equivalent to a MAC channel) is presented. 

To reduce the influence caused by randomness, we take the expectation value over 20 random experiments for each network scenario. 
Basic simulation parameters are listed in Table \ref{tab:network-set}.
Unless otherwise specified, we take the ratio between the minimum and maximum eigenvalues $\eta$ as $4\times10^{-3}$, and we use three moments for the computation, i.e., $N=3$ in \eqref{app_LSD}. 
We perform $15$ iterations of the power method to compute the approximate maximum eigenvalue.
All the experiments are conducted on a platform with an Intel(R) Core(TM) i5-12600KF CPU (10 cores) and 32G RAM.
% The program was locked on a single thread to avoid the influence of multi-thread acceleration.

\begin{table}[htbp]
\vspace{-1em}
\caption{The Network Setting} \label{tab:network-set}
\begin{center}
\begin{tabular}{c c}
\hline
\textbf{Definition and Symbol} & \textbf{Value} \\
\hline
Network scale ($D$)  & 2000m \\
Near field threshold ($d_0$)  & 10m \\
Far field threshold ($d_1$)  & 50m \\
Power limit ($P$)  & 1W \\
Noise power ($N_0$)  & $1\times 10^{-12}$ W \\
Number of clusters ($M$)  & 25 \\
\hline
\end{tabular}
\label{parameters}
\end{center}
\vspace{-1em}
\end{table}

In Fig. \ref{cap_with} (top row), we present three different network scenarios.
\begin{itemize}
    \item Square Uniform: uniformly distributed network nodes in the square network area (left column);
    \item Circle Uniform: uniformly distributed network nodes in the circle network area (middle column);
    \item Circle Normal: truncated normally distributed network nodes in the circle network area (right column);
\end{itemize}
In the bottom row of Fig. \ref{cap_with}, we also present the capacity $C_m$ derived through two different methods (CDM and MPM), under different users-to-BSs ratios $\beta$, and different numbers of BSs $J_m$. 
Note that CDM computes $C_m$ according to the original capacity formula \eqref{cap1} directly, and thus can be considered as the baseline method. 
It can be observed from Fig. \ref{cap_with} that the capacity obtained by our MPM is almost the same as the baseline results. 
This demonstrates the high accuracy of our method. 
For the MU-MIMO systems in Fig. \ref{cap_without}, similar results can be observed. 
Thus, we can draw the same conclusion that our MPM is with high accuracy on capacity estimation. 
In addition, Fig. \ref{cap_with}-\ref{cap_without} also demonstrate the generality of our MPM, since it has high accuracy under different network node distributions, different network shapes, and different values of $\beta$.
% {\color{red}simulation results of $C_m$ obtained by MPM, and numerical results of $C_m$ based on the original formula.} 

%We choose the cluster closest to the center of the network, which is circled in black, to investigate its capacity. (d)-(f) show the capacity derived based on the original formula (blue) and the capacity estimated by MPM (red) with $M=25$. In each figure, the three groups of lines represent $\beta=8,2,0.5$ from top to the bottom. We choose $\eta=0.004$ here.

% Fig. \ref{cap_with} and \ref{cap_without} shows the result of MPM in C$^2$ networks and traditional MU-MIMO cellular networks.
% It is shown that MPM is effective to both C$^2$ networks and MU-MIMO channels.
%In Fig. \ref{cap_with} and \ref{cap_without}, we show the estimation of $C_m$ obtained by MPM based algorithm, {\color{red}and the numerical results of $C_m$ derived based on \eqref{cap1}.} {\color{blue}compared with the capacity computed with \eqref{cap1}.}
%{\color{red}In both figures, two shapes of network areas are considered (i.e., circle and square), and two distributions of network nodes are plotted (i.e. uniform and normal). IS THIS NECESSARY?}
%Specifically, Fig. \ref{cap_with} corresponds to the general C$^2$ architecture, and Fig. \ref{cap_without} corresponds to a special case of C$^2$, where the number of clusters is 1 and equivalent to the MU-MIMO channel. 
%It can be observed from Fig. \ref{cap_with} and \ref{cap_without} that MPM is an accurate estimation on the original expression of the capacity \eqref{cap1} under different network settings. 
%This result means that our algorithm has superior accuracy and generality.

\begin{figure}[htbp]
%\vspace{-1em}
\centerline{\includegraphics[width=9cm]{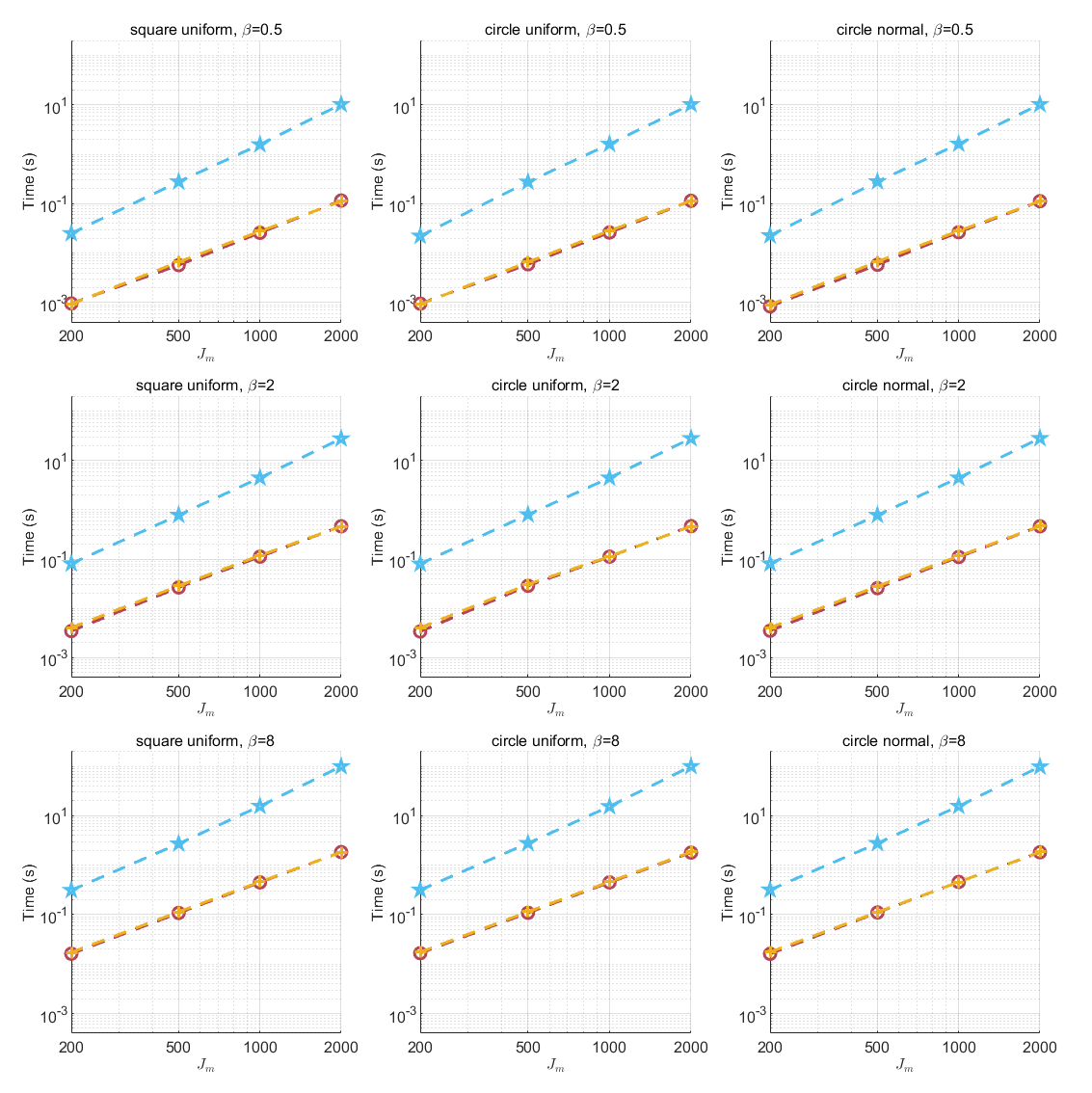}}
\caption{The comparison of computational time of MPM (red circle), TOSE (yellow plus sign) and CDM (blue pentagram). The top row ($\beta=0.5$), middle row ($\beta=2$) and bottom row ($\beta=8$) correspond to different users-to-BSs ratios $\beta$.} %Computational time of MPM compared with TOSE and the traditional Cholesky decomposition based algorithm. We choose $\eta=0.004$, and use 15 iterations in the power method. The blue, red and yellow lines correspond to the baseline algorithm {\color{blue}DO WE NEED TO INTRODUCE THE BASELINE ALGORITHM?}, MPM and TOSE, respectively.}
\label{cap_time}
\vspace{-1.5em}
\end{figure}

To study the efficiency advantage of our MPM, we present the computational time of three different methods in Fig. \ref{cap_time}. 
By data fitting, we can see the empirical complexity of MPM, TOSE and CDM are $O(J_m^{2.05}), \; O(J_m^{2.06})$ and $O(J_m^{2.50})$ respectively. 
Combining the discussions of Fig. \ref{cap_with}-\ref{cap_without}, we can see that although $C_m$ derived by MPM and CDM are almost the same, our MPM is much faster than CDM.

\begin{figure}[htbp]
%\vspace{-1em}
\centerline{\includegraphics[width=9cm]{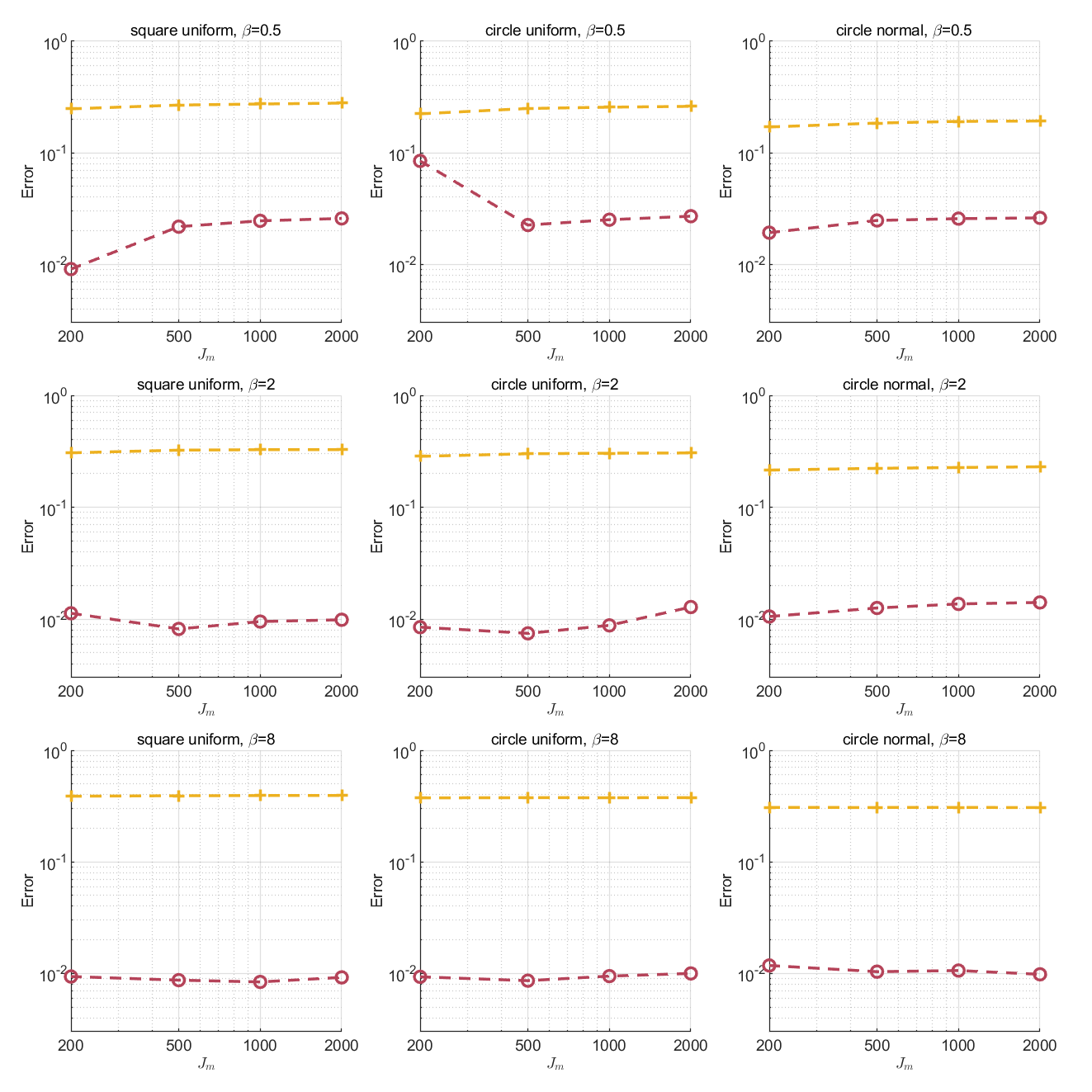}}
\caption{Comparisons of the capacity estimation error between MPM (red circle) and TOSE (yellow plus sign). The top row ($\beta=0.5$), middle row ($\beta=2$) and bottom row ($\beta=8$) correspond to different users-to-BSs ratios $\beta$. }
\label{cap_err}
\vspace{-1em}
\end{figure}

In Fig. \ref{cap_err}, the relative errors of MPM and TOSE are output. 
Here we use CDM as the baseline for comparison. 
It can be seen that MPM is much more accurate than TOSE, although the  computational time of MPM and TOSE are almost the same, as shown in Fig. \ref{cap_time}. 
Thus, we can conclude that MPM is much better than TOSE for capacity estimation. 
In addition, in almost all numerical simulations, the relative error of MPM is less than $2\%$.%, which is sufficient in practice.

%Fig. \ref{cap_time} shows the computational time of MPM compared with our previous algorithm TOSE {\color{red}\cite{jiang2022tose} IS THIS CITATION NECESSARY?} and the traditional Cholesky decomposition based algorithm. 
%It can be seen that the time cost of MPM and TOSE are almost the same, which are both far faster than the traditional Cholesky decomposition based method.
%Specifically, the empirical complexity of the traditional Cholesky decomposition based algorithm, MPM and TOSE are , , respectively.
%{\color{red}Therefore, Fig. \ref{cap_time} not only verifies our theoretical analysis that the complexity of our algorithm is $O(n^{2})$, but also shows that MPM is more efficient than the traditional method.}
%{\color{blue}Therefore, Fig. \ref{cap_time} verifies our theoretical analysis that the complexity of our algorithm is $O(n^{2})$, which is much faster than the traditional method. OR: Therefore, Fig. \ref{cap_time}shows that MPM is more efficient than the traditional method, and verifies our theoretical analysis that the complexity of our algorithm is $O(n^{2})$.}

\begin{figure}[htbp]
%\vspace{-1em}
\centerline{\includegraphics[width=9cm]{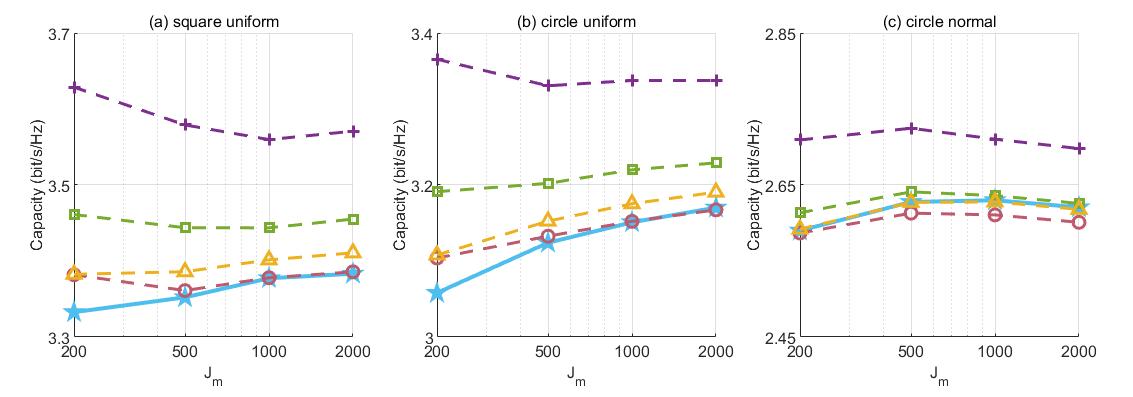}}
\caption{The comparison of MPM with different moments. The baseline is CDM (blue solid line with pentagram). The other four dashed lines show the capacity estimated by using the first one (purple plus sign), two (green square), three (red circle) and four (yellow triangle) moments.
%The comparison of MPM with different moments. The baseline is CDM (blue pentagram). The other four dashed lines show the capacity estimated by using the first one (purple), two (green), three (red) and four (yellow) moments.
}
\label{cap_moments}
\vspace{-1em}
\end{figure}

In \eqref{app_LSD}, we modify the original MP-law with an $N-$th order polynomial to approximate the LSD of the SINR matrix $\mbA_m$. Therefore, a natural question is whether we can obtain better accuracy for capacity as the polynomial order $N$ increases. In Fig. \ref{cap_moments}, the effect of different moments $N$ on the capacity are presented. The results are as we expected. The capacity computed by MPM keeps approaching that of CDM with the increase of $N=1,2,3$. However, the capacity estimation of four moments $N=4$ seems to be slightly worse than that of three moments. This may be due to Runge's phenomenon of high order polynomial interpolation. Thus, $N=3$ of MPM could be a good choice for capacity estimation.

%Moreover, we can see that the capacity of MPM with $N=3$ is accurate enough.
% \footnote{{\color{red}In Fig. \ref{cap_moments}, we find that the capacity estimation results of the three moments are slightly better than that of four moments. This may be due to the inaccurate estimation of the maximum and minimum eigenvalues and the Runge's phenomenon of high order polynomial interpolation. Thus, However, it does not affect our overall conclusion.}} 

% The effectiveness of MPM with different orders of moments. The blue solid lines correspond to {\color{red}the capacity derived from its original formula} {\color{blue} the capacity $C_m$}. The estimated capacity of MPM are plotted with purple, green, yellow and red lines, corresponding to MPM with the first one moment, two moments, three moments and four moments.

%To vividly show the effectiveness of our MPM based algorithm, we plot Fig. \ref{cap_moments}, where we list the result with different orders of moments in our MPM base algorithm.
%{\color{red}Fig. \ref{cap_moments} reflects} {\color{blue} It is shown} that as we take higher orders of moments, the estimation becomes more accurate.
%Then it is reasonable to believe that by taking more orders of moments, the MPM based algorithm is more accurate, while the complexity is controlled at  $O(J_m^2)$.

\begin{figure}[htbp]
%\vspace{-1em}
\centerline{\includegraphics[width=9cm]{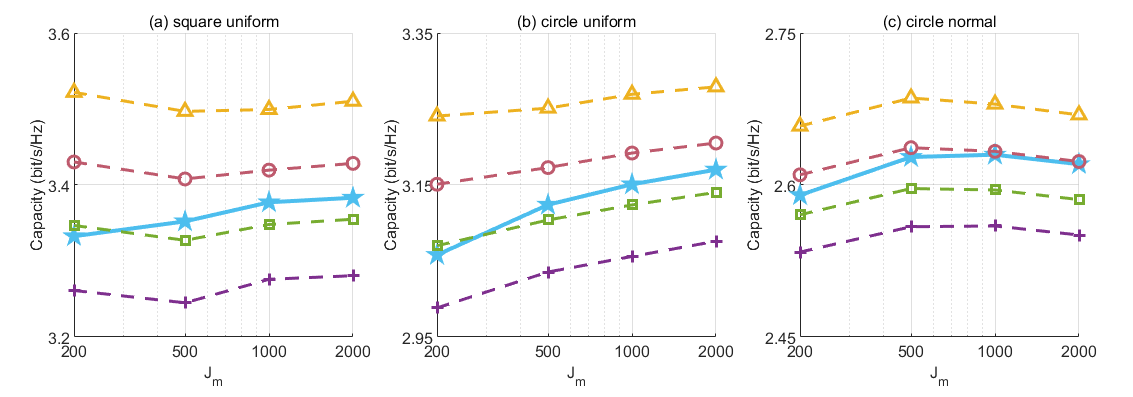}}
\caption{The comparison of MPM with different hyper-parameter $\eta$. The baseline is CDM (blue solid line with pentagram). The other four dashed lines show the capacity estimated with $\eta=0$ (purple plus sign), $\eta= 10^{-3}$ (green square), $\eta=4\times 10^{-3}$ (red circle) and $\eta=10^{-2}$ (yellow triangle).
%The comparison of MPM with different hyper-parameter $\eta$. The baseline is CDM (blue pentagram). The other four dashed lines show the capacity estimated with $\eta=0$ (purple), $\eta=10^{-3}$ (green), $\eta=4\times 10^{-3}$ (red) and $\eta=10^{-2}$ (yellow).
}
\label{cap_eta}
\vspace{-1em}
\end{figure}

Finally, we study the effect of hyper-parameter $\eta$ on the estimation capacity. As discussed at the end of Section 3, computing the non-zero minimum eigenvalue is very difficult and computationally expensive. Therefore, we approximate the maximum-minimum eigenvalue ratio by introducing the hyper-parameter $\eta$. After obtaining the maximum eigenvalue $b$, we can directly estimate the minimum eigenvalue $a=\eta b$. In Fig. \ref{cap_eta}, we present the capacity estimation results under different $\eta$. Meanwhile, the errors of capacity estimation with different $\eta$ are also presented in Tab. \ref{err_eta}. From the figure and table, we can see that the result of capacity estimation is not sensitive to $\eta$. Moreover, $\eta=4\times10^{-3}$ seems to be an optimal choice.

% The effectiveness of our MPM based algorithm with different $\eta$. The blue solid lines correspond to {\color{red}the capacity derived from its original formula} {\color{blue} the capacity $C_m$}. The dotted lines show the result of MPM with $\eta=0$ (red), $\eta=0.001$ (green), $\eta=0.004$ (purple) and $\eta=0.01$ (yellow).

%Fig. \ref{cap_eta} shows the simulation results of estimating the capacity by MPM with different choices of $\eta$.
%{\color{blue}The detailed data is shown in Table \ref{err_eta}.}
%It shows that the estimation results are insensitive to the hyper-parameter $\eta$.
%In most cases, the best choice of $\eta$ is around 0.004.
%The relative error of MPM with different $\eta$ are listed in Table \ref{err_eta}.
%If we always choose $\eta=0$, the relative error is lower than $5\%$.
%The most appropriate $\eta$ is 0 for the case of $\beta=1$, {\color{red}and larger than 0 for other values of $\beta$.} {\color{blue} and gets larger as $\beta$ increases or decreases.}

\begin{table}[htbp]
\vspace{-1em}
\caption{Capacity estimation errors of MPM with different hyper-parameter $\eta$ for circle uniform scenario.}
\vspace{-2.5em}
\begin{center}
\begin{tabular}{c| c c c c c}
\hline
\diagbox{$\eta \;(10^{-3})$}{$\beta$} & $\frac{1}{8}$ & $\frac{1}{2}$ & $2$ & $8$ & $32$ \\
\hline
$0$   & $2.12\%$ & $4.91\%$ & $2.92\%$ & $1.46\%$ & $1.23\%$ \\
$0.5$ & $1.76\%$ & $3.35\%$ & $1.54\%$ & $1.09\%$ & $1.02\%$ \\
$1$ & $1.62\%$ & $2.72\%$ & $1.00\%$ & $0.94\%$ & $0.93\%$ \\
$2$ & $1.44\%$ & $1.85\%$ & $\textbf{0.50\%}$ & $0.72\%$ & $0.80\%$ \\
$4$ & $1.19\%$ & $\textbf{0.68\%}$ & $1.28\%$ & $0.42\%$ & $0.60\%$ \\
$8$ & $0.85\%$ & $0.94\%$ & $2.88\%$ & $\textbf{0.32\%}$ & $0.35\%$ \\
$10$ & $0.72\%$ & $1.53\%$ & $3.53\%$ & $0.48\%$ & $\textbf{0.28\%}$ \\
$20$ & $\textbf{0.27\%}$ & $3.74\%$ & $6.01\%$ & $1.30\%$ & $0.53\%$ \\
% \hline
% calculated & 5.0e-2 & 1.5e-2 & 3.2e-3 & 4.4e-3\\
\hline
\end{tabular}
\label{err_eta}
\end{center}
\vspace{-1.5em}
\end{table}

% \begin{figure}[htbp]
% %\vspace{-1em}
% \centerline{\includegraphics[width=9cm]{net_plot_6.png}}
% \caption{Illustration of three different scenarios of the ultra-dense network. Each color represents an individual cluster in the network. The brighter circles represent BSs and the darker diamonds represent users. We will compute the average capacity of the cluster closest to the center of the network, which is circled in black.}
% \label{net_plot}
% \end{figure}

\section{Conclusion}

In this work, we propose the Moment Passing Method (MPM) to fast and accurately determine the capacity of ultra-dense and complicated networks. 
We can derive the moments of LSD of the SINR matrix with Stieltjes transform and Laurent expansion, which is not affected by different distributions of BSs and users and the shape of network areas. 
As such, we obtain the approximated LSD and the estimated capacity. 
Our MPM is feasible for both C$^2$ networks and conventional MU-MIMO networks, which shows great potential in the network design and analysis of the upcoming B5G/6G era. 

It should be emphasized that the key of this work is the polynomial correction of the classical MP-law. 
Numerically, we can observe nice results for capacity estimation.
It is worth studying the theoretical results of this correction, e.g., the convergence and the applicability to different problems.
We are currently working on this and hope to report the progress in a future paper.

\normalem
\bibliographystyle{IEEEtran}
%\bibliography{cited.bib}

\begin{thebibliography}{99}
% \bibitem{xu2021ten}
% W. Xu, G. Zhang, B. Bai, C. Ai, and J. Wu, “Ten key ICT challenges in the post-Shannon era,” \textit{Scientia Sinica (Mathematica)}, vol. 51, no. 7, pp. 1095–1138, 1 2021.

\bibitem{shannon1948mathematical}
C. E. Shannon, ``A mathematical theory of communication,'' \textit{The Bell system technical journal}, vol. 27, no. 3, pp. 379–423, 1948.

\bibitem{cover2006elements}
T. M. Cover and J. A. Thomas, \textit{Elements of Information Theory}, 2nd ed. John Wiley \& Sons, 2006.

\bibitem{tse2005fundamentals}
D. Tse and P. Viswanath, \textit{Fundamentals of Wireless Communication}. Cambridge University Press, 2005.

% \bibitem{tulino2004random}
% A. M. Tulino, S. Verdú. ``Random matrix theory and wireless communications,'' \textit{Foundations and Trends in Communications and Information Theory}, vol. 1, no. 1, pp. 1--182, 2004.

% \bibitem{telatar1999capacity}
% E. Telatar, ``Capacity of multi-antenna Gaussian channels,'' \textit{European transactions on telecommunications}, vol. 10, no. 6, pp. 585–595, 1999.

\bibitem{yang2021future}
L. Yang, P. Li, M. Dong, B. Bai, D. Zaporozhets, X. Chen, W. Han, and B. Li, ``{C2}: A capacity-centric architecture towards future wireless networking,'' \textit{IEEE Trans. Wireless Commun.}, DOI: 10.1109/TWC.2022.3164286.

\bibitem{wang2021clustered}
J. Wang, L. Dai, L. Yang, and B. Bai,  ``Clustered cell-free networking: a graph partitioning approach,'' 2021, submitted to \textit{IEEE Trans. Wireless Commun.} 2021.

\bibitem{jiang2022tose}
D. Jiang, H. Han, L. Yang, and R. Wang, ``TOSE: A fast capacity determination algorithm based on spike approximations,'' submitted to \textit{IEEE VTC-FALL}, 2022.

\bibitem{tulino2004random}
A. M. Tulino, S. Verd\'u. ``Random matrix theory and wireless communications,'' \textit{Foundations and Trends in Communications and Information Theory}, vol. 1, no. 1, pp. 1--182, 2004.

\bibitem{deng2022cgn}
C. Deng, L. Yang, H. Wu, D. Zaporozhets, M. Dong, and B. Bai, ``CGN: A capacity-guaranteed network architecture for future ultra-dense wireless systems,'' accepted by \textit{IEEE ICC} 2022.

\bibitem{bai2010spectral}
Z. Bai and J. W. Silverstein, \textit{Spectral analysis of large-dimensional random matrices}. Springer, 2010, vol. 20.

\bibitem{wall1948analytic}
H. S. Wall, \textit{Analytic theory of continued fractions}. Van Nostrand, 1948.

\bibitem{rudin1966real}
W. Rudin, \textit{Real and Complex Analysis}. McGraw-Hill, 1966.

\bibitem{hachem2007deterministic}
W. Hachem, P. Loubaton, and J. Najim. ``Deterministic equivalents for certain functionals of large random matrices,'' \textit{The Annals of Applied Probability}, vol. 17, no. 3, pp. 875--930, 2007.

\bibitem{lloyd1982least}
S. P. Lloyd. ``Least squares quantization in {PCM},'' \textit{IEEE Trans. Inf. Theory}, vol. 28, no. 3, pp. 129--137, 1982.

\end{thebibliography}

\end{document}